\documentclass[11pt,a4paper]{article}
\usepackage{graphicx}
\usepackage{jheppub}
\title{Chiral vortices in relativistic hydrodynamics}
\author{Manavendra Mahato}
\affiliation{School of Sciences, Indian Institute of Technology Indore,\\
IET campus, Khandwa Road, Indore, India}
\emailAdd{manav@iiti.ac.in}
\abstract{Towards modelling the charge asymmetry observed in heavy ion collisions, we present here analytic solutions of relativistic hydrodynamics containing parity violating and anomalous terms at the first order in the hydrodynamic approximation. These terms can induce chiral magnetic and chiral vortical effect leading to the generation of the charge asymmetry. We also consider sphaleron solutions with non trivial winding number to model the phenomenon. We calculate the net chiral charge difference produced in our solutions. We anticipate their relevance also in the context of baryogenesis in early universe, neutron star and some condensed matter situations.}
\arxivnumber{1207.0461}
\begin{document}
\maketitle
\flushbottom
 \def\a{\alpha} \def\as{\asymp} \def\ap{\approx}
\def\b{\beta} \def\bp{\bar{\partial}}
 \def\cA{{\cal{A}}}\def\cD{{\cal{D}}} \def\calO{{\cal{O}}} \def\cL{{\cal{L}}} \def\cP{{\cal{P}}} \def\cR{{\cal{R}}}
 \def\da{\dagger} \def\d{\delta}
\def\D{\Delta}
 \def\e{\eta} \def\ep{\epsilon} \def\eq{\equiv}
 \def\f{\frac}
\def\g{\gamma} \def\G{\Gamma}
 \def\hs{\hspace}
\def\i{\iota}
\def\k{\kappa}
\def\lf{\left} \def\l{\lambda} \def\la{\leftarrow} \def\La{\Leftarrow} \def\Lla{\Longleftarrow}\def\lg{\lgroup} \def\Lra{\Longrightarrow} \def\L{\Lambda}
\def\m{\mu}
\def\n{\nu} \def\na{\nabla} \def\nn{\nonumber}
\def\o{\omega}\def\O{\Omega}
\def\p{\phi} \def\P{\Phi} \def\pa{\partial} \def\pr{\prime}
\def\r{\rho} \def\ra{\rightarrow} \def\Ra{\Rightarrow}\def\ri{\right} \def\rg{\rgroup}
\def\s{\sigma} \def\sq{\sqrt} \def\S{\Sigma} \def\si{\simeq} \def\st{\star}
\def\t{\theta}\def\ta{\tilde{\a}} \def\ti{\tilde} \def\tm{\times} \def\tV{\tilde{V}} \def\tr{\textrm} \def\T{\Theta}
 \def\u{\upsilon}
\def\U{\Upsilon}
\def\v{\varepsilon} \def\vh{\varpi} \def\vk{\vec{k}} \def\vp{\varphi} \def\vr{\varrho} \def\vs{\varsigma}\def\vt{\vartheta}
 \def\w{\wedge}
 \def\z{\zeta}
\newcommand{\be}{\begin{equation}} \newcommand{\ee}{\end{equation}}
\newcommand{\bea}{\begin{eqnarray}} \newcommand{\eea}{\end{eqnarray}} 

\begin{section}{Introduction and motivation}
 For the early phase during the formation of quark gluon plasma in heavy ion collisions, relativistic hydrodynamics offers a good model to describe the dynamics. Relativistic hydrodynamics extends the regime of applicability of hydrodynamics to the situations of fast moving fluids such as plasma. Such cases are also encountered in nuclear physics and astrophysics apart from the heavy ion collision experiments. The equations governing the dynamics of the relativistic fluid are conservation equations for the stress energy tensor and the conserved current(s). These are relativistic analogs of the continuity equation and Navier Stokes equations.

 Recently, it was shown that the relativistic hydrodynamics can contain terms which do not have any analogs in non-relativistic cases. The conserved current can contain parity violating terms proportional to vorticity and/or magnetic field. These terms were first noticed during the investigation of equations governing small perturbations at the boundary of the charged black branes and drawing their parallels with Navier-Stokes equations.\cite{Bhattacharyya:2007vs,Bhattacharyya:2008ji,Banerjee:2008th} The theory of gravity in such cases is related to the strong t' Hooft coupling limit of a dual large N field theory and only the long range modes survive in the hydrodynamic approximation, i.e. modes surviving in long temporal and large wavelength limit.  These extra terms were also later understood in terms of quantum triangle anomalies.\cite{Son:2009tf} The anomalies are usually calculated in the perturbative limit of the quantum theory where the coupling is small. The topological nature of anomalies has a role to play to account for their occurrence in both the approaches to the quantum theory. We will discuss here two recently discovered terms in the expression of the current which are also related to anomaly of the previously conserved current. Both of them are parity odd terms. First one is proportional to vorticity  and it leads to the phenomenon of chiral vortical effect. The second term is proportional to the magnetic field and results in chiral magnetic effect.\cite{Fukushima:2008xe}. For non abelian hydrodynamics particularly suited for quark gluon plasma, an effective action for fluid with anomalies was constructed and chiral magnetic effect was obtained from it in \cite{Nair:2011mk}.

 In the case of quark gluon plasma generated during collision of heavy ions, the anomalous current of interest can be axial current leading to net chirality difference along the direction of background magnetic field. The net chirality difference can induce an electric field and a flow of charged carriers along the direction of magnetic field. This is known as chiral magnetic effect.\cite{Fukushima:2008xe} The rate of chiral charge difference depends on the strength of the anomaly. For a fluid with certain configurations for its velocity field ($u^{\m}$), similar buildup of the electric field along the vorticity vector ($\o ^{\m}=\f{1}{2}\ep ^{\m\n\r\s}u_{\n}\pa _{\r}u_{\s}$) happens during the process of chiral vortical effect. Such processes are not only restricted to quark gluon plasma, but are also important in many other contexts (to be discussed below). In all these cases, relativistic hydrodynamics offers itself as a good candidate model to describe the dynamical process.

 In quark gluon plasma, the gluon field can locally have some topologically non trivial configurations. These configurations are classified by their winding numbers and are separated from each other by a potential barrier of the order of $\L _{\tr{QCD}}$. The system can tunnel through it due to an instanton. But such processes are exponentially suppressed at high densities and at weak couplings.\cite{Son:2004tq} The system can also roll over the barrier and such transitions are called sphalerons which manifest themselves during the presence of anomalies. These metastable configurations are quantum fluctuations possibly of the order of few fms and lead  to a difference in left and right handed quarks. The discrete symmetries such as parity and CP are locally violated in these spatially localized bubbles.  A sphaleron in non-abelian gauge fields in QCD can offer an explanation to the strong CP problem. They can also be present in quark-gluon plasma in a high magnetic field.\cite{Oz:2011am, Son:2007ny} Thus, they can lead to an observable charge separation by generating the net chiral difference and subsequent interaction of chiral particles with background magnetic field leading to chiral magnetic effect.

  The sphaleron offers a mechanism to explain many different effects in different settings. For example, sphaleron processes can be partly responsible for the baryon asymmetry in our universe. Before electroweak symmetry breaking, a non trivial topological configuration of hypercharge electromagnetic field could have occurred in the early universe plasma. Such a sphaleron configuration offers a model to generate the requisite amount of baryon asymmetry needed for the subsequent process of nucleosynthesis.\cite{Giovannini:1997eg} These processes violate baryon number conservation as well as local P and CP symmetries. The rate of baryon asymmetry production was estimated in \cite{Kuzmin:1985mm} and it was found that such explanations are favourable if the electroweak symmetry breaking in the early universe was of first order. Recently, early universe baryogenesis due to anomaly in lepton number current was also considered.\cite{Dvornikov:2011ey}

The anomalies in 4 dimensional field theories can also be understood in terms of their dual gravity solutions (if they exist) in the context of  AdS/CFT  correspondence.\cite{Maldacena:1997re, Witten:1998qj, Newman:2005hd, Landsteiner:2011cp, Kalaydzhyan:2011vx} The well understood \cal{N}=4 Super Yang Mills theory can have anomalous R charge currents. It maps to a nontrivial 5 dimensional bulk Chern-Simons term in the gravitational Lagrangian. The magnetic field acting as a source in the chiral magnetic effect amounts to putting appropriate boundary conditions on the bulk electromagnetic tensor $F_{\m\n}$ in the dual gravity solution. The anomalous current will correspond to another abelian gauge field in the bulk. Holographic computations were also performed to determine chiral magnetic effect for the case of anisotropic fluid and its dependence on elliptic flow coefficient $\u _2$, a quantity which parameterizes the event charge anisotropy in heavy ion collisions.\cite{Gahramanov:2012wz}.

  Superfluids can also display a similar phenomena called chiral electric effect.\cite{Bhattacharya:2011tra, Neiman:2011mj} Here, an anomalous current is generated due to a topologically non-trivial configuration of electric field. Moreover, chiral vortical effect in a pionic superfluid leads to a flow of fermionic zero modes along the direction of vorticity.\cite{Kirilin:2012mw} For chiral magnetic effect in the same medium, there will be strings carrying magnetic flux and the magnetic field plays the role of the vorticity vector. Chiral magnetic effect can also be present in metal crystals having a non trivial Berry phase configuration in the presence of electromagnetic field.\cite{Son:2012wh} If there are $k$ quanta of Berry curvature flux associated with any given Fermi surface, then the fermionic number current will be anomalous with the anomaly proportional to $k \vec{E}.\vec{B}$. This triangle anomaly will also give its contribution to the density-density correlator. If there are 2 Fermi surfaces with unequal chemical potentials in the presence of magnetic field, the chiral magnetic effect will manifest itself as a flow of fermionic current between the Fermi surfaces with a strength proportional to the magnetic field and the difference in chemical potentials.

 In this paper, we will explore these parity violating effects in the case of quark gluon plasma using the hydrodynamic approach. We will partly develop and demonstrate two methods of constructing  hydrodynamic solutions containing these parity violating and anomalous terms. We will keep the dissipative coefficients vanishing throughout, so our fluid can be interpreted as a perfect fluid in the presence of parity violating and anomalous coefficients. In quark gluon plasma, the viscosity is actually very small.\cite{Teaney:2009qa} In section 2, we write down the equations for the relativistic hydrodynamics at the first order explaining our conventions. Some solutions witnessing chiral vortical effect can be constructed by first finding a relativistic generalization of some known non-relativistic solutions and then modifying them to admit the non-trivial vorticity terms. This is demonstrated in section 3 for the case of a famous non relativistic solution known as Taylor-Green vortex. We will keep the electromagnetic fields vanishing here and calculate the net axial charge difference generated which amounts to chiral vortical effect. The second method is to use Hopf mapping to construct a topological solution with winding number one using the velocity and electromagnetic fields. A non relativistic magnetohydrodynamic solution based on such mapping was constructed earlier in \cite{kamchat}. However, we find that Hopf mapping can be used to generate a larger set of solutions of relativistic hydrodynamics, potentially setting a stage to explore in detail many dynamic processes.  This will be the content of section 4. We will also find some simple solutions using this method in this section. In section 5, we use the same method to generate a sphaleron solution in relativistic hydrodynamics with a topologically non trivial configuration of the background electromagnetic field. This solution has all the parity violating and anomalous coefficients non trivial at the first order. Even though, topological configurations for non-abelian fields seem to be more interesting, we will restrict ourselves to U(1) fields only in this paper for simplicity. In the case of quark gluon plasma, it can be thought of as a restricted abelian version of chromo-electromagnetic fields or a background U(1) field produced by the highly energetic colliding charged particles. Also, all the quantities denoting space-time dimensions are kept dimensionless throughout in this paper. The point of view is to assume some natural length scale present in the theory and then dimensionless position variables denote multiples of it. The natural length scale in the case of quark gluon plasma can be inverse of either dynamically generated QCD scale, $\L _{\tr{QCD}}$ or center of mass energy, $s_{\tr{cm}}$ or temperature, $T$. We consider our solutions as representative solutions displaying chiral effects and calculate chiral charge difference contributed by them. Furthermore, we relax the boundary conditions to simplify the construction of the analytic solutions. In many cases of potential applications as mentioned above, boundary conditions may not play any significant role in determining the underlying processes. We also didn't impose the equation of state and it is determined implicitly by the additional assumptions that we make to simplify our equations. We consider physically interesting cases to be those for which the pressure and energy density are both positive everywhere.
\end{section}

\begin{section}{Equations for anomalous hydrodynamics}
The equations for relativistic hydrodynamics are given as conservation equations for the stress-energy tensor, $T^{\m\n}$ and the conserved current(s). See eqs. (\ref{hydroequations}). We will consider the case of one conserved current, $j^{\m}$ which can represent particle flux like baryon current or axial current and is likely to get anomalous contributions. We also have background electromagnetic fields with the electric and magnetic fields defined as $E^{\m}=F^{\m\n}u_{\n}$ and $B^{\m}=\f{1}{2}\ep ^{\m\n\a\b}u_{\n}F_{\a\b}$, where $F_{\m\n}$ is an antisymmetric field strength tensor. In the fluid rest frame, their spatial components indeed represent electric and magnetic fields. The conservation equations are supplemented with constituent equations which express stress energy tensor($T^{\m\n}$) and current($j^{\m}$) in terms of pressure($P$) , enthalpy density($h$), temperature ($T$), particle number density ($n$), velocities ($u^{\m}$) and their derivatives. Fluid is likely to have non trivial vorticity defined as $\o ^{\m}=\f{1}{2}\ep ^{\m\n\r\s}u_{\n}\pa _{\r}u_{\s}$ giving rise to a parity violating term in the current conservation equation. In the presence of electromagnetic fields, one can have a parity violating term proportional to magnetic field as well as an anomalous term in the current equation. The presence of such terms was also understood in terms of constraints on near equilibrium partition function.\cite{Banerjee:2012iz}. They were also calculated using Kubo formula and do receive corrections proportional to gravitational anomaly coefficient.\cite{Landsteiner:2011cp, Landsteiner:2011iq}.
  We write below equations for relativistic hydrodynamics containing these terms.
 \bea
 \label{heT}
 \pa _{\m}T^{\m\n}&=&F^{\n\l}j_{\l}\nn\\
 \label{heaj}
 \pa _{\m}j^{\m}&=&-\f{C}{8}\ep ^{\m\n\r\s}F_{\m\n}F_{\r\s}=CE^{\m}B_{\m}\nn\\
 \label{hect}
 T^{\m\n}&=&hu^{\m}u^{\n}+Pg^{\m\n}-\eta P^{\m\a}P^{\n\b}(\pa _{\a}u_{\b}+\pa _{\b}u_{\a})-\lf (\z -\f{2}{3}\eta\ri )P^{\m\n}\pa _{\l}u^{\l}\nn\\
 \label{hydroequations}
 j^{\m}&=&nu^{\m}-\s TP^{\m\n}\pa _{\n}\lf (\f{\m}{T}\ri )+\s E^{\m}+\xi \o ^{\m}+\xi _B B^{\m}
 \eea
 These are supplemented with the relativistic constraint on velocity, $u^{\m}u_{\m}=-1$.  Here, $\mu$ is the chemical potential and  $g^{\m\n} $ is the metric, which we take to be Lorentzian with signature $(-,+,+,+)$. The notation, $P_{\m\n}=g_{\m\n}+u_{\m}u_{\n}$ denotes the projection tensor and projects any tensor perpendicular to the velocity field. It is a symmetric tensor and satisfies relation $u_{\m}P^{\m\n}=0$. The density $n$ may represent axial charge density in the case of quark gluon plasma or baryon/lepton charge density in the case of baryogenesis in the early universe. The various dissipative coefficients are bulk viscosity ($\z$), shear viscosity ($\eta$) and  conductivity ($\s$). We will call the coefficients of parity violating terms $\xi$ and $\xi _B$ as chiral vortical conductivity and chiral magnetic conductivity, respectively. The coefficient $C$ denotes the strength of the anomaly. It is related to chiral conductivities as
\bea
\xi &=&C\lf (\mu ^2-\f{2}{3}\f{n\mu ^3}{\ep +P}\ri )\nn\\
\xi _B&=&C\lf (\mu-\f{1}{2}\f{n\mu ^2}{\ep +P}\ri )
\label{ccreln}
\eea
 The electromagnetic fields also satisfy Maxwell equations i.e. field strength conservation equation and Bianchi identity.
 \bea
 \pa _{\mu}F^{\m\n}&=&j^{\n}_{\tr{EM}},\nn\\
 \pa_{[\m}F_{\n\r ]}&=&0.
 \eea
 The quantity $j^{\m}_{EM}$ represents the background electromagnetic four-current i.e $j^{0}_{EM}$ represents the charge density and its spatial part, $j^{i}_{EM}$ represent the electric current, where $i=1,2,3$. The convention for Levi-Civita symbol used in this manuscript is $\ep _{0123}=\ep^{+-12}=1$.
 \end{section}
\begin{section}{Solution by uplifting}
In this section, we will try to estimate the chiral effects due to a fluctuation plausible in quark gluon plasma. To do so, we will first construct an explicit relativistic hydrodynamic solution by uplifting a solution of non-relativistic hydrodynamics. Even if we expect that the observable quantities in quark gluon plasma may not strongly depend on the details of such fluctuations, construction of few explicit representative solutions will make possible any explicit computations of relevant quantities. Here, we choose a well known solution of non-relativistic hydrodynamics called Taylor-Green solution to estimate its chiral effects.
The non-relativistic equations are given as
\bea
\f{\pa v^1}{\pa x}+\f{\pa v^2}{\pa y}=0\nn\\
\f{\pa v^1}{\pa t}+v^1\f{\pa v^1}{\pa x}+v^2\f{\pa v^1}{\pa y}&=&-\f{1}{\r _n}\f{\pa P_n}{\pa x}+\nu\lf (\f{\pa ^2 v^1}{\pa x^2}+\f{\pa ^2 v^1}{\pa y^2}\ri )\nn\\
\f{\pa v^2}{\pa t}+v^1\f{\pa v^2}{\pa x}+v^2\f{\pa v^2}{\pa y}&=&-\f{1}{\r _n}\f{\pa P_n}{\pa y}+\nu\lf (\f{\pa ^2 v^2}{\pa x^2}+\f{\pa ^2 v^2}{\pa y^2}\ri )
\eea
Here, $v^i \;(i=1,2),\; P_n,\;\ep _{n} ,$ and $\rho _{n}$ denote velocity components, pressure, energy density and mass density of a non-relativistic fluid in 2+1 dimensions. They are continuity equation and conservation of momentum flux for constant mass density $\rho _{n}$.
 We then try to modify their solutions to also accommodate new coefficients. The first part can be done using the prescription given in \cite{Rangamani:2008gi,Rangamani:2009xk}. According to it, given a solution of non-relativistic equations in 2+1 dimensions, a solution of relativistic equations in $3+1$ dimensions can be written as
\bea
u^+&=&\sqrt{\f{1}{2}\f{\rho _{n}}{\ep _{n}+P_{n}}}\nn\\
u^-&=&\f{1}{3}\lf (\f{1}{u^+}+u^+v^2\ri )\nn\\
u^i&=&u^+v^i\nn\\
P&=&P_{n}\nn\\
\label{NrtR}
\r&=&2\ep _{n}+\rho_{n}
\eea
Here, $u^+$ and $u^-$ denote the velocity components along the null directions i.e. $u^{\pm}=\f{1}{\sqrt{2}}(u^0\pm u^z)$ and $u^i$, the same along other two spatial directions. The coordinates along the three spatial directions are denoted as $x,\; y$, and $z$. Symbols $P$ and $\r$ denote relativistic pressure and density of the fluid.

The Taylor Green vortex solution is given as in \cite{NSE},
\bea
v^1 &=& F(t)\sin x\cos y,\nn\\
v^2&=& -F(t)\cos x\sin y,\nn\\
F(t)&=& e^{-2\nu t},\nn\\
P_n&=&\f{\r _n}{4}F(t)(\cos 2x+\cos 2y).
\eea
We consider the simple case of zero viscosity. We put $\nu =0$ here and then try to get the relativistic version of it. This solution has a relativistic analog in 3+1 dimensions as discussed in eqs. (\ref{NrtR}).
\bea
u^+&=& \lf[2\f{\ep _n}{\r _n}+\f{1}{2}(\cos 2x+\cos 2y)\ri]^{-1/2},\nn\\
u^-&=&\f{u^+}{2}\lf [1+\f{2\ep _n}{\r _n}-2\sin ^2x\sin ^2y\ri ],\nn\\
u^x&=&u^+ \sin x\cos y,\nn\\
u^y&=&-u^+ \cos x\sin y,\nn\\
P&=&\f{\r _n}{4}(\cos 2x+\cos 2y),\nn\\
\label{relEQ}
h&=&\f{\r_n}{(u^+)^2}.
\eea
The above quantities satisfy the relativistic equation for perfect fluid i.e.
\be
\pa _{\mu}T^{\mu\nu}=0.
\ee
To find the modified solution of relativistic equations with non-trivial new coefficients, we first simplify the relativistic hydrodynamic equations. We choose the first order dissipative coefficients $\z =\s=\eta =0$  and write $F^{\m\n}$ in terms of electric and magnetic fields as
\be
F_{\m\n}=2u_{[\m}E_{\n ]}-\ep _{\m\n\r\s}u^{\r}B^{\s}
\ee
The conservation equations are then
\bea
\pa _{\m}T^{\m\n}&=&nE^{\n}+u^{\n}[\xi _{B}E^{\m}B_{\m}+{\xi}E_{\r}\o ^{\r}-\xi u_{\r}z^{\r}]-\xi z^{\n}\nn\\
\pa_{\m}j^{\m}&=&CE^{\m}B_{\m}
\eea
where
\be
z^{\m}=B_{\s}\pa ^{[\mu}u^{\s ]}
\ee
We then simplify the equations by taking the background electromagnetic fields to be vanishing, i.e. $E_{\m}=B_{\m}=0$. We further break the equation with stress tensor into its trace and traceless part by taking the trace with the velocity vector, $u^{\m}$. After a little algebraic manipulation, we get the following equations,
\bea
n\pa _{\mu}(h u^{\mu})-h j^{\r} (u. \pa) u_{\r}&=&(j.\pa)P,\nn\\
\pa . j&=&0,\nn\\
hu^{\m}\pa _{\m}u^{\n}+u^{\n}u^{\m}\pa _{\m}P+\pa ^{\n}P&=&0.
\label{HydroEqn}
\eea
We find that the relativistic analog constructed in eqs. (\ref{relEQ}) helps a lot in dealing with these equations. To accommodate the non trivial vorticity term with chiral vortical conductivity $\xi$, it suffices to choose a suitable expression for the number density $n$.  We present our solution in terms of Lorentz factor $\chi$, which in this case turns out to be
\be
\chi^{-2}=\f{2\ep _n}{\r _n}+\f{1}{2}(\cos 2x +\cos 2y)
\ee
 This variable is equivalent to $u^+$ in eqs. (\ref{relEQ}). The solution can then be written as
\bea
h&=&\f{\r _n}{\chi^2 },\nn\\
u^{\mu}&=&\chi v^{\mu},\nn\\
v^{+}&=&1,\nn\\
v^{-}&=&\f{1}{2}+\f{\ep _n}{\r _n}-\sin ^2x\sin ^2y,\nn\\
v^{1}&=&\sin x\cos y,\nn\\
v^{2}&=&-\cos x\sin y,\nn\\
P&=&P_0+\f{\r _n}{4}(\cos 2x+\cos 2y),\nn\\
n&=&\f{\xi}{3}\chi\sin x\sin y.
\label{SSoln}
\eea
We consider $P_0$ to be a constant positive quantity. The chiral vortical conductivity appears explicitly in the last expression for the number density $n$. The only connection between the two conservation equations is the velocity. The current is
\bea
j^{-}&=&\f{\xi}{3}\chi^2\sin x\sin y\lf (1+\f{2\ep _n}{\r _n}-\sin ^2 x -\sin ^2 y\ri ),\nn\\
j^ +&=&j^x=j^y=0.
\eea
This solution is in a steady state. This is expected as we have dropped all dissipative terms. The non-trivial chiral vortical conductivity $\xi$, does not lead to dissipation. We calculate the zeroth component of vorticity to be
 \be
 \o ^0=\f{\chi ^2}{3\sqrt{2}}\sin x\sin y\lf[ \f{\ep _n}{\r _n}-\f{3}{2}+\cos ^2x\cos ^2y\ri].
 \ee
 The contribution to axial charge difference will be $\int d^3x \xi\o ^0$. We assume that the solution holds for some length $L$ along the z direction. Along other two spatial directions, the contributions from different regions tend to cancel due to sinusoidal dependence. So, the net contribution depends on how far the solution extends. The maximum contribution will be for the situation in which the solution extends from $0$ to $\pi$ along both x and y directions.
 Then the axial charge difference generated will be
 \begin{figure}
\begin{center}
\includegraphics[width=7cm,clip]{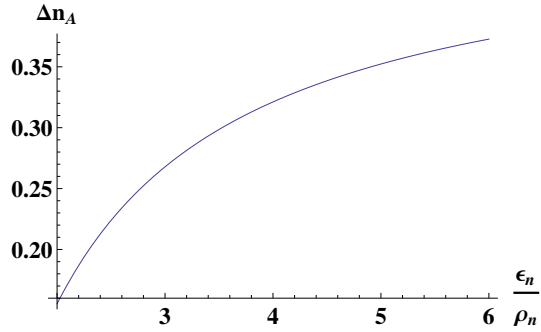}
\caption[Short caption for figure 1]{\label{AxialCD} Plot of maximum axial charge separation (in units of $\xi L$) as a function of ratio of energy and number density, $\ep _n/\rho _n$ generated for the case of relativistic Taylor Green vortex solution. }
\end{center}
\end{figure}
 \be
 \D n_A=\f{\xi L}{3\sqrt{2}}\int_0^{\pi} dx\int_0^{\pi} dy \f{\sin x\sin y(\ep _n/\r _n-3/2+\cos ^2x\cos^2y)}{(2\ep _n/\r_n-1+\cos ^2x+\cos^2y)}.
 \ee
 The axial charge difference created will induce an electric field resulting in chiral vortical effect. It is evaluated numerically and displayed as a function of $\ep _n/\rho _n$ in fig. (\ref{AxialCD}).
 \end{section}
\begin{section}{Solution using Hopf fibration}
The background electromagnetic fields generated in quark gluon plasma are due to charges of the colliding ions. Due to the high energy present in the collisions, a non trivial configuration of the background field is also likely to occur.
 In this section, we try to find relativistic solutions with topologically non trivial background electromagnetic field. An interesting non-relativistic solution is given in \cite{kamchat}, which has a non-trivial index defined as
\be
I=\f{16}{\pi ^2}\int\vec{A}.(\na \times \vec{A})d^3x.
\ee
The solution was obtained there using Hopf fibration map.
Given such a map $f: S^3\ra S^2$, one can pullback the volume form of 2-sphere to get a two form on 3-sphere. Any two form on a 2-sphere can be written as an exact differential of a one form. The vector dual of such a one form can be projected on ${R}^3$ using stereographic projection. The value of the above index is one for the vector potential in $R^3$ generated by this procedure.
The latter is given as \cite{kamchat}
\bea
A^1&=&\f{(xz-y)}{2r^2},\nn\\
A^2&=&\f{(yz+x)}{2r^2},\nn\\
A^3&=&\f{(2z^2+2-r)}{4r^2},\nn\\
{\tr{where }}r&=&1+x^2+y^2+z^2.
\eea
We will use this vector potential to generate solutions of relativistic hydrodynamic equations which are more general than the simple relativistic generalization of the solution in  \cite{kamchat}.
We make the assumption that the profiles of the fluid $(u^{\m})$ and that of the background vector potential $(\hat{A}^{\m})$ are proportional to the vector potential given above, i.e.,
\bea
u^{\m}&=&(v, fA^i),\nn\\
\hat{A}^{\m}&=& (\b ,\a A^i).
\eea
Here, $f$, $v$, $\b$ and $\a$ are functions of the radial coordinate $r=1+x^2+y^2+z^2$ and  time $t$.
The electro-magnetic tensor is calculated to be
\bea
F^{0i}&={\pa ^{0}}{\hat{A}^i}-{\pa ^{i}}{\hat{A}^0} &= -(\dot{\a}A_i+2\b ' x_i),\nn\\
F^{ij}&={\pa ^{i}}{\hat{A}^j}-{\pa ^{j}}{\hat{A}^i} &= \ep ^{ijk}\lf [2\lf (\f{2\a}{r}-\a '\ri)A_k+\f{r\a '}{2}a_k\ri ],\nn\\
&{\tr{where}}\quad\quad a_i&= \f{1}{r^2}(-y,x,1).
\eea
We denote the derivatives with respect to time and radial coordinate $r$ by dot $\dot{}$ and prime $'$, respectively. The Bianchi identity is automatically satisfied by this construction. We get the following expressions for the electric and magnetic fields.
\bea
E^0&=&-f\lf (\f{\dot{\a}}{16r^2}+\f{z\b '}{2r}\ri ),\nn\\
E^i&=&-A_i\lf (v\dot{\a}+\f{fz\a '}{2r} \ri )+x_i\lf (-2v\b ' +\f{f\a '}{8r^2}\ri ),\nn\\
B^0&=&-\f{f\a}{12r^3},\nn\\
B^i&=&-\f{2}{3}\lf (\f{2v\a}{r}-v\a '+\b 'f\ri )A_i+\f{ra_i}{6}(\b 'f-v\a ').
\eea
The vorticity here is
\bea
\o ^0&=&-\f{f^2}{24r^3},\nn\\
\o^i&=&\f{A_i}{3}\lf (\f{-2vf}{r}+vf'-v'f\ri )+\f{ra_i}{12}(v'f-vf').
\eea
The equations that need to be satisfied are the conservation equation for stress energy tensor and the current equation along with the relativistic constraint on velocity.
\bea
\pa _{\m}T^{\m\n}&=&F^{\m\l}j_{\l},\nn\\
\pa _{\m}j^{\m}&=&CE^{\m}B_{\m},\nn\\
T^{\m\n}&=&hu^{\m}u^{\n}+P\eta ^{\m\n}\nn\\
j^{\mu}&=& n u^{\m}+\xi \o ^{\m}+\xi _BB^{\m},\nn\\
u^{\m}u_{\m}&=&-1.
\eea
Here, we have taken the conductivity and viscosities to be zero $(\s=\eta=\z=0)$.
These equations for our case reduce to
\bea
A^i\lf [ (vhf\dot{)}\ri . &+&\lf . \f{z}{2r^2}(rhf^2)'+nv\dot{\a}\ri ]
+x_i\lf ( 2P'-\f{hf^2}{8r^3}+2nv\b '\ri )\nn\\
&=&\lf (\f{z}{r^2}A^i-\f{x_i}{4r^3}\ri )\lf [-\f{rnf\a '}{2}+\f{\xi}{3}\{(\a v)'f-\a vf' \}+\f{2}{3}\xi _B\a f\b '\ri ]+\nn\\
&&+\f{f}{24r^3}(\dot{\a}A^i+2\b 'x_i)(\xi f+2\xi _B\a),\nn\\
(hv^2\dot{)}&+&\f{z}{2r^2}(rvhf)'-\dot{P}\nn\\
&=&\f{1}{48r^3}(\dot{\a }+8r\b 'z)(-3rnf+2\xi vf+4\xi _B v\a ),\nn\\
(nv\dot{)}&+&\f{z}{2r^2}(rnf)' -\xi \f{f\dot{f}}{12r^3}-\f{2\xi}{3r^2}zfv '\nn\\
&=&\f{\xi _B}{12r^3}(f\a \dot{)}+\f{2z\xi _B}{3r^2}(v'\a+\b 'f)+\f{\a C}{12r^3}(\dot{\a}+8rz\b '),\nn\\
v^2&=&1+\f{f^2}{16r^2}.
\eea
 They are coupled non-linear partial differential equations and in order to solve them, we make an ansatz that the factors proportional to $A^i$, $zA^i$, $x^i$ and $z$ cancel out separately. The resulting equations can be written elegantly in terms of two new variables defined as
\bea
M&=&-rnf+\f{2}{3}v(\xi f+2\xi _B\a),\nn\\
N&=&nv-\f{f}{24r^3}(\xi f+2\xi _B\a)=-\f{v}{rf}M+\f{2}{3rf}(\xi f+2\xi _B\a).
\eea
We thus obtain the following set of equations.
\bea
(vhf\dot{)}+N\dot{\a}&=&0,\nn\\
\a 'M+\f{2\xi}{3}\a (fv'-vf')+\f{4\xi _B}{3}\a(f\b '-v\a ')&=&(rhf^2)',\nn\\
P'+\f{(hf^2)'}{16r^2}+\b 'N&=&0,\nn\\
\dot{P}-(hv^2\dot{)}+\f{\dot{\a}}{16r^3}M&=&0,\nn\\
\dot{N}&=&\f{C\a\dot{\a}}{12r^3},\nn\\
(rvhf)'&=&\b 'M,\nn\\
\label{MasterEqn}
M'+\f{2\xi}{3}(fv'-vf')+\f{4\xi_B}{3}(f\b '-v\a ')+\f{4}{3}C\a \b '&=&0.
\eea
We next choose Coulomb gauge $\hat{A}^0=\b =0$ without loss of generality and for simplicity, we look for only steady state solutions. In other words, we assume that all the functions explicitly appearing in the set of equations above are time-independent. We denote $rvhf$ by a constant $k$. This reduces the above set of equations to
\bea
P'&=&-\f{1}{16r^2}(hf^2)'\nn\\
\label{eqnN}
\lf (\f{k f}{v}\ri )'+(rnf)\a '&=&\f{2\xi}{3} [(\a v)'f-(\a v)f']\\
\label{eNp}
(rnf)'&=&\f{4}{3}v'(\xi f+\xi _B\a).
\eea
Here, $v$ and $f$ are related by the constraint $v^2=1+\f{f^2}{16r^2}$. The equations, even with the strident looking assumptions, admit a wide class of solutions. We consider two cases.

{\bf{Case I:}} $C=0$.

This is a perfect fluid case with no chiral vortical and chiral magnetic conductivity and is the simplest non trivial solution of this class. The eq. (\ref{eNp}) tells that $rnf=N_0$ (a constant). The other equation (\ref{eqnN}) can then be solved to obtain the following solution.
\bea
\a &=&-\f{4k rf}{N_0\sqrt{16r^2+f^2}},\nn\\
h&=&\f{4k}{f\sqrt{16r^2+f^2}},\nn\\
n&=&\f{N_0}{rf},\nn\\
P&=&P_0-4k\int _{1}^r \f{\r}{(16\r ^2+f(\r)^2)^{3/2}}\lf (\f{f(\r)}{\r}\ri)' d\r.
\eea
Here, $P_0$ represents the pressure at the origin, $x=y=z=0$. The function $f(r)$ is left undetermined and can be any function which leads to well defined physical quantities i.e. pressure, energy density and enthalpy density.

{\bf{Case II:}}  $v=\l f$, $\l$ is a constant.

In this case, the spatial part of vorticity is proportional to spatial part of velocity or background vector potential. Velocity constraint and equation for pressure leads to
\bea
f&=&\f{4r}{\sqrt{16\l ^2r^2-1}},\nn\\
P&=&\f{k}{48\l}\lf (1-\f{1}{r^3}\ri )+P_0.
\eea
Here, $P_0$ is a positive constant denoting the pressure at the origin.
We find the number density using equation (\ref{eqnN}) to be
\be
n=\f{2\xi \l}{3}\f{f}{r}
\ee
The fact that $rvhf=k$ can be used to evaluate the enthalpy density.
\be
h=\f{k}{\l rf^2}.
\ee
Equation (\ref{eNp}) leads to a consistency relation stating that $\xi _B\a =0$ which implies $\a =0$. This is then a case of pure gauge and hence no electromagnetic fields.
 One gets the 4-vector $\o ^{\mu}$ and $j^{\m}$ for this case to be
\bea
\o ^{\m}&=&\lf (-\f{f^2}{24r^3},-\f{2\l f^2}{3r}A^i \ri ),\nn\\
j^{\m}&=&\lf(\f{2\xi}{3r},0,0,0\ri )
\eea

Near the origin $x=y=z=0$ i.e. $r=1$, the solution simplifies to
\bea
u^{\m}&=&\f{4}{\sqrt{16\l ^2-1}}\lf [\l,0,0,\f{1}{4}\ri ],\nn\\
n&=&\f{8\xi \l}{3\sqrt{16\l ^2-1}},\nn\\
h&=&\f{k(16\l ^2-1)}{16\l},\nn\\
P&=&P_0.
\eea
To see the asymptotic behavior of the solution at large distances $r\ra \infty$, we use spherical coordinates $\t ,\p $, so that $x=\sqrt{r-1}\sin \t\cos\p ,y=\sqrt{r-1}\sin\t\sin\p $ and $z=\sqrt{r-1}\cos\t$. Then as $r\ra \infty$,
\bea
u^{\m}&\ra & \lf( 1, \f{1}{4\l r}\sin (2\t)\cos\p ,\f{1}{4\l r}\sin (2\t)\sin\p, \f{1}{4\l r}\cos(2\t)\ri ),\nn\\
n&\ra &\f{2\xi}{3r},\nn\\
h&\ra &\f{k\l}{r},\nn\\
P&\ra &P_0+\f{k}{48\l}.
\eea
From the pressure profile, we see that such configurations can happen around local depressions in pressure. As $r$ increases, the pressure increases to a constant value. The fluid velocity, number density and enthalpy decreases as $1/r$. The magnitude of velocity or speed is spherically symmetric at large $r$. This solution suffers a problem that energy density becomes negative for r larger than $r\sim 48\l ^2(1+48k\l /P_0)^{-1}$. So, solution can be relevant only for small $r$ and should be dominated by some other solution at larger $r$. It can happen that some of the dissipative coefficients neglected here can prevent the solution from this problem. To calculate the contribution to the axial charge difference caused by the vorticity, we assume the solution to hold upto $r=R$. We thus obtain
\bea
\D n _A&=&\int d^3x \xi \o ^0=-\f{4\pi\xi}{3}\int _1^R\f{\sqrt{r-1}dr}{r(16\l ^2r^2-1)}\nn\\
&=&\f{2\pi\xi}{3\sqrt{\l}}\lf [\sqrt{4\l-1}\tan ^{-1}\lf (2\sqrt{\f{\l(R-1)}{4\l -1}}\ri )
+\sqrt{4\l+1}\tan ^{-1}\lf(2\sqrt{\f{\l(R-1)}{4\l+1}}\ri )\ri ]-
\nn\\&&
-\f{8\pi\xi}{3}\sec^{-1}\sqrt{R}
\eea
This axial charge difference will induce an electric current and hence, will lead to chiral vortical effect.
\end{section}
\begin{section}{Sphaleron Solution}
We revisit the eqs. (\ref{MasterEqn}) again and  simplify them by choosing $v=\l f$ and $\b= \l \a$. We also choose $\a=k f$. These assumptions make the spatial part of vorticity, magnetic field, velocity and vector potential proportional to each other. We look here for steady state solutions only. Then, the set of equations needed to be solved is
\bea
(rhf^2)'&=&\a 'M,\nn\\
M&=&-\f{2}{3}\l C\a ^2+a_0,\nn\\
P'+\f{(hf^2)'}{16r^2}+\l \a 'N &=&0.
\eea
We solve top two equations above to obtain
\bea
f&=&\f{4r}{\sqrt{16\l ^2r^2-1}},\nn\\
h&=&-\f{2}{9}\l Ck^3\f{f}{r}+a_0 k\f{1}{rf}+\f{b_0}{rf^2},\nn\\
n&=&\f{2\l f}{3r}(Ck^2+\xi +2k\xi _B)-\f{a_0}{rf}.
\eea
We simplify the equation for pressure to get
\be
P'=\f{8}{3}\l k \lf (\xi +2k\xi_B+\f{2}{3}Ck^2\ri )\f{1}{r(16\l ^2r^2-1)^{3/2}}+\f{b_0}{16r^4}.
\ee
This results in the expression for pressure as
\be
P=P_0+\f{8}{3}\l k \lf (\xi +2k\xi_B+\f{2}{3}Ck^2\ri )\lf (\tan^{-1}\f{f}{4r}-\f{f}{4r}\ri )-\f{b_0}{48r^3}.
\ee
As $r\ra \infty$, the asymptotic behavior of various expressions above are
\bea
h&\ra &\f{k^2}{r}\lf (a_0+b_0-\f{2}{9}C\l\ri ),\nn\\
n&\ra &\lf [\f{2\l}{3k}(Ck^2+\xi +2k\xi _B)-a_0 k\ri ]\f{1}{r},\nn\\
P&\ra &P_0-\f{1}{48r^2}\lf [b_0+\f{2\l}{9k^2}\{2Ck^2+3(\xi+2k\xi_B)\}\ri ].
\eea
The associated background electric and magnetic fields are calculated to be
\bea
E^{\m}&=&\lf ( \f{\l kf^4}{32r^4}z, \f{kf^4}{32r^4}zA^i+\f{kf^2}{8r^3}x^i\ri ),\nn\\
B^{\m}&=&\lf (-\f{kf^2}{12r^3},-\f{4\l k}{3r}f^2A^i\ri ).
\eea
The vorticity and the anomalous current are
\bea
\o^{\m}&=&\lf (-\f{f^2}{24r^3},-\f{2\l}{3r}f^2A_i\ri ),\nn\\
j^{\m}&=&\lf [\f{2}{3r}(Ck^2\l ^2f^2+\xi+2k\xi _B)-\f{a_0 \l}{r} ,\lf (\f{2C\l k^2}{3}\f{f^2}{r}-\f{a_0}{r}\ri )A^i\ri ].
\eea
The background electromagnetic current is found to be
\bea
j^{0}_{\tr{EM}}&=&2\l k \{2(r-1)f''+3f'\}=\f{3\l kf^5}{128r^5}\{1+16\l ^2r(r-2)\},\nn\\
j^i_{\tr{EM}}&=&2kA^i\lf \{-\f{4}{r}\lf(\f{2f}{r}-f'\ri)-2f'+2\lf(\f{2f}{r}-f'\ri)'+(rf')'\ri\}+\nn\\
&&+kra^i\lf(\f{2f}{r}-f'\ri)'-\f{k(rf')'}{2r}\d ^i_z\nn\\
&=&-\f{kf^5A^i}{128r^6}\lf [12\lf(\f{4r}{f}\ri)^4+2(7-2r)\lf(\f{4r}{f}\ri)^2+3(2-r)\ri]-\nn\\
&&-\f{k\l ^2f^5a^i}{16r^3}(1+32\l ^2r^2)-\f{kf^5}{2^9r^6}(1+32\l ^2r^2)\d ^i_z.
\eea
There are 6 unknown parameters, namely $\l, k, C, a_0, b_0, P_0 $ along with 2 chiral conductivities $\xi$ and $\xi_B$. Parameter $\l$ is related to the ratio of fluid velocity to speed of light (usually denoted as $\b$) as $\l =(4r \b)^{-1}$. Parameters $k$ and $C$ denote the strength of the electromagnetic field and the anomaly, respectively. Constants $a_0$, $b_0$ and $P_0$ are related to the axial charge density, enthalpy and pressure in the absence of non-trivial topological charge for this solution. We take these constants to be zero as we are interested in effects due to topological charge. We can then simplify the chiral conductivities  related by eq. (\ref{ccreln}) to obtain
\bea
\xi &=&\f{C\mu ^2}{1-2\mu/k}\nn\\
\xi _B&=&\f{C\mu(1-\mu/(2k))}{1-2\mu/k}
\eea

 Real values of $f$ needs $\l >1/4$.  We consider physically interesting cases as those for which $P$ and $\ep =h-P$ are always positive. All possible choices of parameters do not lead to physically sensible results. Below, we present one such choice, though other choices can also be feasible. In quark gluon plasma, fluid velocity can be very high up to fractions of speed of light.\cite{KerenZur:2010zw} We choose parameter $\l =1$ which implies that the maximum fluid velocity in the configuration is a quarter of the speed of light. Quantity $C$, related to anomaly in
 the corresponding field theory , should be some number of order unity. We choose it to be $C=1$. Strength of the electromagnetic field and chemical potential vary a lot from one region in space-time to other.\cite{Fukushima:2008xe} We choose $k$ related to electromagnetic field to be $-1$. Minus sign helps us to get physically sensible results for this configuration. We choose $\mu=1/2$ for the same reason. It is also a typical value of chemical potential for strong magnetic field.\cite{Fukushima:2008xe} For this choice, the radial dependence of enthalpy density and pressure as well as the implied equation of state are plotted in figures \ref{PlotPHE} and \ref{EquationState}.
\begin{figure}
\begin{minipage}[t]{7cm}
\begin{center}
\includegraphics[width=7cm,clip]{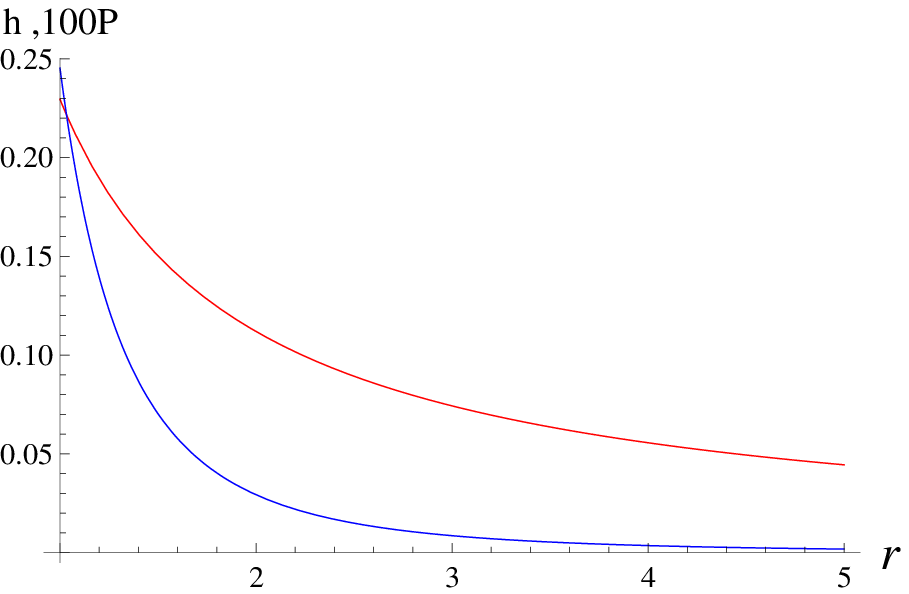}
\caption[Short caption for figure 1]{\label{PlotPHE} Plot of enthalpy density ({\it{red curve}}) and 100 times the pressure ({\it{blue curve}}) vs $r$. The maximum of these physical variables occur at the origin.}
\end{center}
\end{minipage}
\hspace{0.5 cm}
\begin{minipage}[t]{7cm}
\begin{center}
\includegraphics[width=7cm,clip]{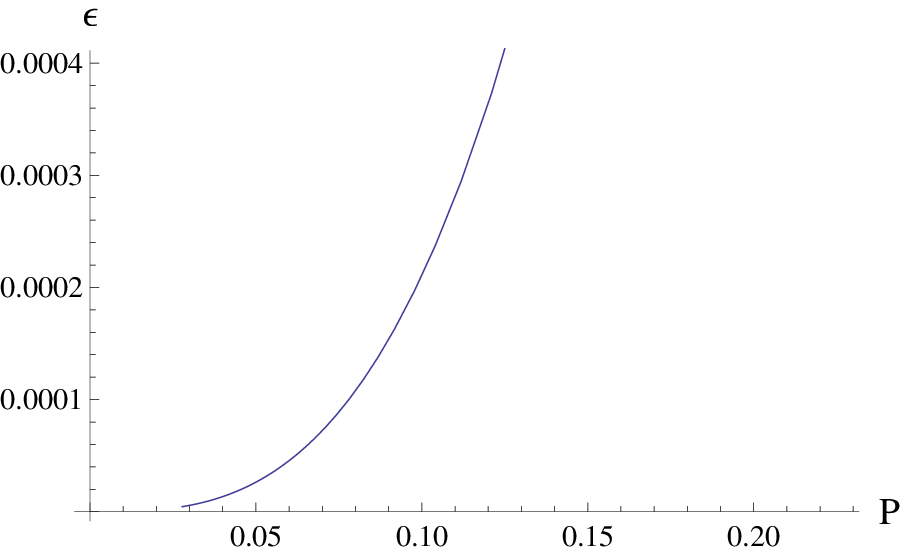}
\caption[Short caption for figure 2]{\label{EquationState} Energy Density is plotted against pressure. The energy density is more sensitive to changes in pressure at higher pressure, i.e. near the origin.}
\end{center}
\end{minipage}
\end{figure}

The anomaly present in the right hand side of eq. (\ref{heaj}) i.e. $CE^{\m}B_{\m}$, can be written as a total derivative of a Chern Simons current given as
\be
K^{\m}=-\f{C}{4}\ep ^{\m\n\r\s}\hat{A}_{\n}F_{\r\s}.
\ee
The Chern Simons charge of this sphaleron is
\bea
N_{CS}&=&\int d^3xK^0=4C\pi k^2\int _1^{\infty}\f{\sqrt{r-1}dr}{r(16\l ^2r^2-1)}\nn\\
&=&-\f{C\pi ^2k^2}{\sqrt{\l }}(\sqrt{4\l +1}+\sqrt{4\l -1}-4\sqrt{\l})
\eea
The rate of topological winding number changing transitions caused by the sphaleron solution will be proportional to the above charge, $N_{CS}$. In the case of quark gluon plasma, it will contribute towards to the rate of production of chirality difference and the induced electric current generated. However, the chiral charge difference will also get contributions due to non-trivial vorticity and background magnetic field. Since, $j^{\m}=nu^{\m}+\xi \o^{\m}+\xi_B B^{\m}$,
the contribution due to both vorticity and magnetic field is
\bea
\D \tilde{n}&=&-\int d^3x (\xi \o ^0+\xi_B B^{0} )\nn\\
&=&\f{(\xi+2k\xi_B)}{24}\int \f{f^2}{r^3}d^3x\nn\\
&=&-\f{\pi ^2(\xi+2k\xi_B)}{3\sqrt{\l}}(\sqrt{4\l +1}+\sqrt{4\l -1}-4\sqrt{\l}).
\eea
Hence, the total chiral charge difference created is
\be
\D n_A=-\f{\pi ^2(3Ck^2+\xi+2k\xi_B)}{3\sqrt{\l}}(\sqrt{4\l +1}+\sqrt{4\l -1}-4\sqrt{\l}).
\ee
This chiral charge difference will induce an electric field resulting in a combination of chiral magnetic effect and chiral vortical effect. In the case of plasma in early universe, the above expression will be proportional to the rate of baryogenesis.
\end{section}

\begin{section}{Conclusion and future directions}
 Chiral magnetic and vortical effect due to sphalerons has evinced much interest recently because it is a candidate model for explaining the observed charge dependent azimuthal asymmetries in the heavy ion collisions.\cite{Kharzeev:2009fn} We attempted to model the charge asymmetry generation process in quark gluon plasma by constructing in this work explicit, analytic solutions of relativistic hydrodynamics containing parity violating and anomalous terms. Some solutions constructed in sections 3 and 4 are devoid of electromagnetic fields, though they possess non-trivial vorticity. These can nevertheless be candidate model for processes involving chiral vortical effect. The sphaleron solution constructed in section 5 is richer and have non-trivial values for all the parity violating and anomalous terms possible at the first order in the hydrodynamic approximation. It displays a combination of both chiral vortical and chiral magnetic effect. Azimuthal correlations between 3 particles are used to measure charge separation effects in experiments. They have been shown to be related to correlators of coefficients of sine terms in the Fourier decomposition of the charged particle azimuthal distribution.\cite{Abelev:2009ac} Such sine terms are manifestly parity violating. The conjectured formation of metastable topological configurations is in line with observed differences between same charge and opposite charge correlators. These effects disappear at low energies which is consistent with the sphaleron model.\cite{Kharzeev:2011vv}  The equations of hydrodynamics admit many different solutions and it is a priori difficult to say  whether a particular solution actually occurs or not. However, the charge separation can be due to many such sphalerons occurring in different point of space and time in the quark gluon plasma. Further observations are needed to fully ascertain whether these parity violating effects can be attributed to such configurations. In this direction, a detailed study of such solutions may facilitate in figuring out some typical predictions out of them and in providing theoretical predictions/hints for future tests.

We believe our solutions also to be relevant in many other different contexts like plasma in early universe, superfluids and Fermi liquids as discussed in the introduction. Another interesting application area of our solutions can be neutron stars. The core of a neutron star is made up of highly dense quark matter displaying superfluid properties.\cite{Baym,Pines,Hsu:1999rf} We anticipate that our results will also be relevant in  describing some properties of vortices in neutron star.
It will be interesting to construct explicit models based on these solutions to quantitatively demonstrate and evaluate the significance of the chiral magnetic and chiral vortical effect in various situations. It will help us to chart out the kinetics and dynamics of many processes in greater detail.

 We will also like to see the role played by dissipative coefficients in these solutions and in the processes in general. Quark gluon plasma possesses negligible viscosity and it is considered nearly perfect fluid. However, inclusion of dissipation in these solutions will enable us to make more realistic models and make better quantitative predictions out of them. Finally, from fluid dynamic point of view, the set of equations given by eqs. (\ref{MasterEqn}) is one of the main outcomes of our analysis. These equations can be used to generate a variety of solutions of relativistic hydrodynamics relevant in different contexts. It is also an interesting problem to investigate that what kind of fluids i.e. fluids specified by equation of state can accommodate hydrodynamic solutions governed by the reduced set of equations written in eq. (\ref{MasterEqn}). We will also like to check the stability and linear response of our solutions against various perturbations. Also, we have only demonstrated that steady state solutions of eq. (\ref{MasterEqn}) are available, but it will be interesting to construct their explicit time dependent solutions or to demonstrate the consistency of these equations including the temporal derivative terms. Nevertheless, our solutions being explicit analytic solutions, can also be used as benchmarks to check numerical codes for relativistic hydrodynamics. Analytic solutions of non-relativistic hydrodynamics have long been used for such purposes.
\end{section}
\acknowledgments
I am grateful to Prof. Spenta Wadia for suggesting me this problem. I thank Spenta Wadia, Sandip Trivedi, Shiraz Minwalla, Ajay Pratap Singh and especially Jyotirmoy Bhattacharya for various comments and discussions. I also like to thank people of India for their support towards research in basic sciences.
\bibliographystyle{jhep}
\bibliography{ahydro}{}

 \end{document}